\documentstyle[twocolumn,prl,aps,epsfig]{revtex}
\begin{document}
\topmargin=0.5cm    
\voffset -0.5in
\draft
\title{The reaction dynamics of the  $^{16}$O$(e,e^{\prime}p)$  cross section  
at  high missing energies} 
\author{
N.~Liyanage,$^{17}$ 
B.~D.~Anderson,$^{13}$
K.~A.~Aniol,$^{2}$
L.~Auerbach,$^{29}$
F.~T.~Baker,$^{7}$
J.~Berthot,$^{1}$
W.~Bertozzi,$^{17}$
P.~-Y.~Bertin,$^{1}$
L.~Bimbot,$^{22}$
W.~U.~Boeglin,$^{5}$
% ...
}
\address{
{\rm (The Jefferson Lab Hall A Collaboration)}\\
\vspace{0.1cm}
$^{1}$Universit\'e Blaise Pascal/IN2P3, F-63177 Aubi\`ere, France\\
$^{2}$California State University, Los Angeles, California 90032, USA\\
% ...
$^{5}$Florida International University, Miami, Florida 33199, USA\\
% ...
$^{7}$University of Georgia, Athens, Georgia 30602, USA\\
% ...
$^{13}$Kent State University, Kent, Ohio 44242, USA\\
% ...
$^{17}$Massachusetts Institute of Technology, Cambridge, Massachusetts 02139, 
USA\\
% ...
$^{22}$Institut de Physique Nucl\'eaire, F-91406 Orsay, France\\
% ...
$^{29}$Temple University, Philadelphia, Pennsylvania 19122, USA\\
% ...
}
\date{\today\nobreak\vspace*{-1.5cm}}

\maketitle

\begin{abstract}
We measured the  cross section and response functions
($R_L, R_T$, and $R_{LT}$) for the $^{16}$O$(e,e^{\prime}p)$ reaction in 
quasielastic kinematics  for missing energies $25 \le
E_{\rm miss} \le 120$ MeV at various missing momenta $P_{\rm miss} \le 340$ 
MeV/$c$.  For
$25 < E_{\rm miss} < 50$ MeV and $P_{\rm miss} \approx 60$ MeV/$c$, the reaction
is dominated by single-nucleon knockout from the 1$s_{1/2}$-state.  At
larger $P_{\rm miss}$, the single-particle aspects are increasingly masked
by more complicated  processes.  For $E_{\rm miss} > 60$ MeV and
$P_{\rm miss} > 200$ MeV/$c$, the cross section is relatively constant. 
Calculations which include contributions from pion exchange currents, isobar 
currents and short-range correlations account for the shape and the 
transversity  but only for half of the magnitude of the measured cross 
section.  
\end{abstract}
\pacs{PACS numbers: 25.30.Fj, 27.20.+n}

The $(e,e^{\prime}p)$ reaction in quasielastic kinematics ($\omega
\approx Q^2/2m_p$)\footnote{The kinematical quantities are: the
electron scattered at angle $\theta_e$ transfers momentum $\vec q$ and
energy $\omega$ with $Q^2 = \vec q\thinspace^2 - \omega^2$.  The
ejected proton has mass $m_p$, momentum $\vec p_p$, energy $E_p$, and
kinetic energy $T_p$.  The cross section is typically measured as a
function of missing energy $E_{\rm miss} = \omega - T_p - T_{\rm recoil}$ and
missing momentum $P_{\rm miss} = \vert \vec q - \vec p_p \vert$.  The polar
angle between the ejected proton and virtual photon is $\theta_{pq}$
and the azimuthal angle is $\phi$.  $\theta_{pq} > 0^\circ$ corresponds
to $\phi = 180^\circ$ and $\theta_p > \theta_q$.  $\theta_{pq} < 0^\circ$
corresponds to $\phi = 0^\circ$.}  has long been a useful tool for the
study of nuclear structure.  $(e,e^{\prime}p)$ cross section
measurements have provided both a wealth of information on the wave
function of protons inside the nucleus and stringent tests of nuclear
theories. Response function measurements have provided detailed
information about the different reaction mechanisms contributing to
the cross section.

In the first Born approximation, the unpolarized 
$(e,e^{\prime}p)$ cross section can be separated into four
independent response functions, $R_{L}$ (longitudinal), $R_{T}$
(transverse), $R_{LT}$ (longitudinal-transverse), and $R_{TT}$
(transverse-transverse)~\cite{kelly1}. These response functions 
contain all the information that can be extracted
from the hadronic system using the $(e,e^{\prime}p)$ reaction.

Originally, the quasielastic cross section was attributed entirely to
single-particle knockout from the valence states of the nucleus.
However, a series of $^{12}$C$(e,e^{\prime}p)$ experiments performed
at MIT-Bates~\cite{ulmer1,lourie1,weinstein1,morrison,holtrop} measured much
larger cross sections at high missing energy  than were expected
by single-particle knockout models.  $^{12}$C$(e,e^{\prime}p)$
response function data reported by Ulmer {\it et al.}\cite{ulmer1}
show a substantial increase in the transverse-longitudinal difference,  
($S_T-S_L$),\footnote{$S_X =
\frac{\sigma_{\rm Mott}V_{\rm X} {R_{\rm X}}}{\sigma^{ep}_X},$ 
where X $\epsilon$ \{$T,L$\}, and $\sigma^{ep}_X$ is calculated from
the off-shell ep cross section obtained using deForest's cc1
prescription~\cite{deforest1,deforest2}.} above the two-nucleon emission 
threshold.  
Similar $R_T/R_L$
enhancement has also been observed by Lanen {\it et al.} for $^6$Li
\cite{lanen}, by van der Steenhoven {\it et al.} for $^{12}$C
\cite{steenhoven1} and, more recently, by Dutta {\it et al.} 
 for $^{12}$C, $^{56}$Fe, and $^{197}$Au~\cite{dutta}.

There have been several theoretical attempts
\cite{ryckebusch,gil,takaki} to explain the continuum strength using
two-body knockout models and final-state interactions, but no single
model has been able to explain all the data.  

In this first Jefferson Lab Hall A experiment~\cite{saha}, we studied the
$^{16}$O$(e,e^{\prime}p)$ reaction in the quasielastic region at $Q^2
= 0.8$ (GeV/$c$)$^2$ and $\omega = 439$ MeV ($|\vec{q}\thinspace| \approx 
1$ GeV/c).  We extracted the $R_L$,
$R_T$, and $R_{LT}$ response functions from cross sections measured at
several beam energies, electron angles, and proton angles for 
$P_{\rm miss} \le 340$ MeV/$c$.   This paper reports the
results for $E_{\rm miss} > 25$ MeV; $p$-shell knock-out region ($E_{\rm miss}
 < 20$ MeV) results from this experiment were reported in \cite{Gao00}.

We scattered the $\sim$70 $\mu$A continuous electron beam from
a waterfall target \cite{gari} with three foils, each $\sim$130
mg/cm$^2$ thick. We detected the scattered electrons and knocked-out protons
in the two High Resolution Spectrometers (HRS$_e$ and HRS$_h$).  The
details of the Hall A experimental setup are given
in~\cite{halla,nilanga}.

We measured the $^{16}$O$(e,e^{\prime}p)$ cross section at three beam
energies, keeping $\vert\vec q\thinspace\vert$ and $\omega$ fixed in
order to separate response functions and understand systematic
uncertainties. Table I shows the experimental kinematics.  

The accuracy of a response-function separation depends on precisely
matching the values of $|\vec{q}\hspace*{0.5mm}|$ and $\omega$ for
different kinematic settings.  In order to match
$|\vec{q}\hspace*{0.5mm}|$, we measured $^1$H$(e,ep)$ (also using the
waterfall target) with a pinhole collimator in front of the HRS$_e$.
The momentum of the detected protons was thus equal to $\vec{q}$.  We
determined the $^1$H$(e,ep)$ momentum peak to $\frac{\delta p}{p} =
1.5\times 10^{-4}$, allowing us to match 
$\frac{\delta | \vec{q}\thinspace|}{|\vec{q}\thinspace|}\hspace*{0.5mm}$ to
$1.5\times 10^{-4}$ between the different kinematic settings.
Throughout the experiment, $^1$H$(e,e)$ data, measured simultaneously with
$^{16}$O$(e,e^{\prime}p)$, provided a continuous monitor of both
luminosity and beam energy.

The radiative corrections to the measured cross sections were
performed by two independent methods; using the code 
RADCOR~\cite{nilanga,quint}, which unfolds the radiative tails in 
$(E_{\rm miss}, P_{\rm miss})$ space, and using the code 
MCEEP~\cite{ulmer2} which simulates the radiative tail based on the  
prescription of Borie and Drechsel~\cite{borie}. The corrected cross 
sections from the two methods agreed within the statistical uncertainties of 
these data. The radiative correction to the continuum cross section for 
$60 < E_{\rm miss} < 120$ MeV was about 10\% of the measured cross section. 

At $\theta_{pq}=\pm 8^{\circ}$, $R_{LT}$ extracted independently
at beam energies of 1.643 GeV and 2.442 GeV agree well within
statistical uncertainties. This indicates that the systematic
uncertainties are smaller than the statistical
uncertainties. The systematic uncertainty in cross section measurements is 
about 5\%. This uncertainty is dominated by the uncertainty in the 
$^1$H$(e,e)$ cross section to which the data were normalized~\cite{simon}.

Figure 1 shows the  measured cross section as a function of missing
energy at $E_{\rm beam} = 2.4$ GeV for various proton angles, $2.5^{\circ} \le
\theta_{pq} \le 20^{\circ}$. The average missing momentum increases with 
$\theta_{pq}$ from 50 MeV/$c$ to 340 MeV/$c$.  The prominent peaks at 12 MeV 
and 18 MeV are due to
$1p$-shell proton knockout and are described in \cite{Gao00}, where it was 
shown that the $p$-shell cross sections can be explained up to 
$P_{\rm miss} = 340$ MeV/$c$ by relativistic Distorted Wave Impulse 
Approximation (DWIA) calculations.  However the spectra for 
$E_{\rm miss} > 20$ MeV exhibit a very different behavior. At the lowest 
missing momentum, $P_{\rm miss} \approx 50$ MeV/$c$, the wide
peak centered at $E_{\rm miss} \approx 40$ MeV is  due predominantly to 
knockout of
protons from the $1s_{1/2}$-state.  This peak is less prominent at
$P_{\rm miss} \approx 145$ MeV/$c$ and has vanished beneath a flat
background for $P_{\rm miss} \ge 200$ MeV/$c$.  At $E_{\rm miss} > 60$ MeV or
$P_{\rm miss} > 200$ MeV/$c$, the cross section does not depend on
$E_{\rm miss}$ and decreases only weakly with $P_{\rm miss}$.

\begin{figure}[H]
\centering \epsfig{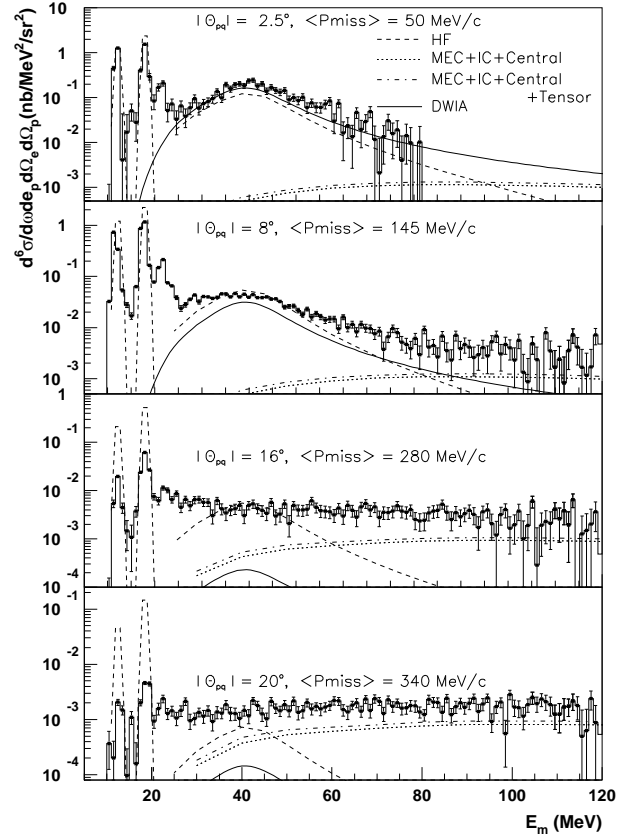} 
\caption 
{Cross sections measured at different outgoing proton angles as a
function of missing energy. The curves show the 
single-particle strength calculated by Kelly (solid curve, only s-shell is 
shown) and by
Ryckebusch (dashed curve), folded with the Lorentzian parameterization
of Mahaux. The dotted line shows the Ryckebusch {\it et al.}
calculations of the $(e,e^{\prime}pn)$ and $(e,e^{\prime}pp)$
contributions to $(e,e^{\prime}p)$ including meson-exchange currents
(MEC), intermediate $\Delta$ creation (IC) and central correlations,
while the dot-dashed line also includes tensor correlations.}
\label{cs}
\end{figure}

We compared our results to single-particle knockout calculations by
Kelly~\cite{kelly2} and
Ryckebusch~\cite{ryckebusch2,ryckebusch3,ryckebusch4} to determine how
much of the observed continuum ($E_{\rm miss} > 20$ MeV) cross section can
be explained by 1$s_{1/2}$-state knockout.  Kelly \cite{kelly2}
performed DWIA calculations using a relativized Schr\"odinger equation
in which the dynamical enhancement of lower components of Dirac
spinors is represented by an effective current
operator \cite{hedayati}. 
These calculations accurately
describe the $1p$-shell missing momentum distributions up to 340 MeV/$c$
\cite{Gao00}.  For the
$1s_{1/2}$-state, Kelly used a normalization factor of 0.73 and spread
the cross section and the response functions over missing energy using the 
Lorentzian
parameterization of Mahaux \cite{mahaux}.  At small $P_{\rm miss}$, where
there is a clear peak at 40 MeV, this model describes the data well.
At larger $P_{\rm miss}$, where there is no peak at 40 MeV, the DWIA cross
section is much smaller than the measured cross section (see
Figure~\ref{cs}).  Relativistic DWIA calculations by other authors
\cite{Picklesimer85,udias} show similar results.  This confirms the
attribution of the large missing momentum cross section to
non-single-nucleon knockout.

Figure~\ref{cs} also shows calculations by Ryckebusch {\it et al.}
~\cite{ryckebusch2,ryckebusch3,ryckebusch4} using a non-relativistic
single-nucleon knockout Hartree-Fock (HF) model which
uses the same potential for both the ejectile and bound nucleons.
Unlike DWIA, this approach conserves current at the one-body level,
but it also requires much smaller normalization factors because it
lacks a mechanism for diversion of flux from the single-nucleon
knockout channel.  At small missing momentum, this model describes
both the $p$-shell and $s$-shell cross sections well.  As the missing
momentum increases, it progressively overestimates the $p$-shell and $s$-shell 
cross
sections.    The most important difference
between the DWIA and HF single-nucleon knockout models is the
absorptive potential; its omission from the HF model increases the HF cross 
section for
$P_{\rm miss} \approx 300$ MeV/$c$ by an order of magnitude for both $p$-shell
and $s$-shell.

Figure~\ref{cs} also shows $(e,e^{\prime}pn)$ and $(e,e^{\prime}pp)$
contributions to the $(e,e^{\prime}p)$ cross section calculated by Ryckebusch 
{\it etal.}~\cite{ryckebusch6}. This calculation has also been performed in a 
HF framework. The cross section for the two particle knock-out has been 
calculated in the ``spectator approximation'' assuming that the two 
knocked-out nucleons will escape from the residual  $A-2$ system without 
being subject to inelastic collisions with other nucleons.  This calculation 
includes contributions mediated by pion-exchange currents, intermediate 
$\Delta$ creation and
central and tensor short-range correlations. According to this calculation, 
in our kinematics, two-body currents (pion-exchange and $\Delta$) account for 
approximately 85\% of the calculated $(e,e^{\prime}pn)$ and $(e,e^{\prime}pp)$ 
strength.  Short-range tensor correlations contribute approximately 13\% while 
short-range central correlations contribute only about 2\%. Since the two-body 
current contributions are predominantly transverse, the calculated  
$(e,e^{\prime}pn)$ and $(e,e^{\prime}pp)$ cross section is mainly transverse 
in our kinematics.  The flat cross section
predicted by this calculation for $E_{\rm miss} > 50$ MeV is consistent
with the data, but it accounts for only about  half the measured cross section.
 Hence, additional contributions to the cross section such as heavier meson 
exchange and processes involving more than two hadrons must be considered.

\begin{figure}[H]
\centering \epsfig{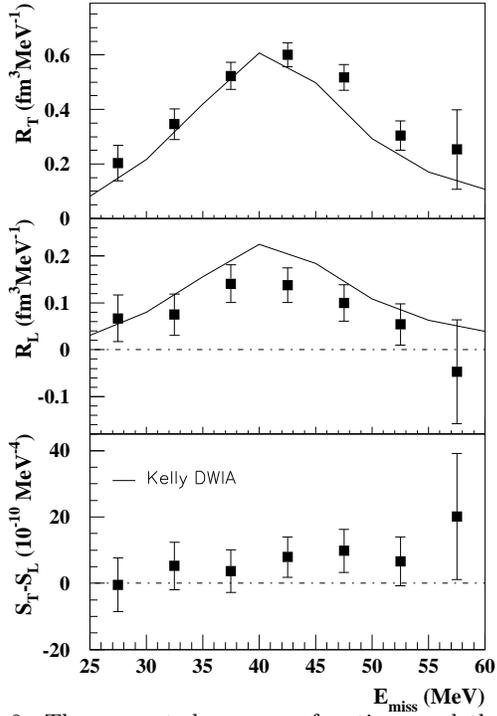}
\caption 
{The separated response functions and the difference of the
longitudinal and transverse spectral functions for 
$\langle P_{\rm miss}\rangle\approx $ 60 MeV/$c$. The calculations have been 
folded with the Lorentzian parameterization
of Mahaux and have been  binned in the same manner as the data.}
\label{responce_par} 
\end{figure}

Figures~\ref{responce_par}-\ref{responce_280} present the separated response 
functions for various proton angles. Due to  kinematic constraints, we were 
only able to  separate the responses for $E_{\rm miss} <60$ MeV.  The separated
 response
functions can be used to check the reaction mechanism.  If the excess
continuum strength at high $P_{\rm miss}$ is dominated by two body
processes rather than by correlations, then it should be predominantly 
transverse.

Figure~\ref{responce_par} presents the separated response functions for
$\langle P_{\rm miss}\rangle
\approx 60$ MeV/$c$.  The wide peak centered around $E_{\rm miss} \approx
40$ MeV in both $R_L$ and $R_T$ corresponds primarily to
single-particle knockout from the 1$s_{1/2}$-state.  The difference
between the transverse and longitudinal spectral functions
($S_T-S_L$), which is expected to be
zero for a free nucleon,  appears to increase slightly with $E_{\rm miss}$.
The magnitude of ($S_T-S_L$) measured here is consistent with the
decrease in ($S_T-S_L$) with $Q^2$ seen in the measurements of Ulmer
{\it et al.}~\cite{ulmer1} at $Q^2 = 0.14$ (GeV/$c$)$^2$ and by Dutta
{\it et al.}~\cite{dutta} at $Q^2 = 0.6$ and 1.8 (GeV/$c$)$^2$. This suggests 
that,
in parallel kinematics, transverse non-single-nucleon knockout
processes decrease with $Q^2$.

Figure~\ref{responce_145} presents the separated response functions 
($R_{L+TT}$\footnote{$R_{L+TT} \equiv R_{L}+\frac{V_{TT}}{V_L}R_{TT}$}, $R_T$, 
 and $R_{LT}$) 
for $|\theta_{pq}| = 8^{\circ}$ ($\langle P_{\rm miss}\rangle \approx 145$
MeV/$c$).  The Mahaux parameterization does not reproduce the shape of
$R_L$ or of $R_T$ as a function of missing energy. For $E_{\rm miss}< 40$
MeV, all calculated response functions underestimate the data suggesting the 
excitation of states with a complex structure between the $p$- and $s$-shells.
  For
$ E_{\rm miss} > 50$ MeV, $R_{L+TT}$ (which is mainly longitudinal because
$\frac{V_{TT}}{V_L}R_{TT}$ is estimated to be only about 7\% of
$R_{L}$~\cite{kelly2} in these kinematics) is consistent with both zero and 
with the
calculations.  $R_T$, on the other hand, remains nonzero to at least
60 MeV.   $R_T$ is also significantly larger than the DWIA
calculation.  $R_{LT}$ is about twice as large as the DWIA
calculation over the entire range of $E_{\rm miss}$.  $R_{LT}$ is nonzero
for $E_{\rm miss} > 50$ MeV, indicating that $R_L$ is also nonzero in that
range.

Figure~\ref{responce_280} presents the separated response functions
for $|\theta_{pq}|=16^{\circ}$ ($\langle P_{\rm miss}\rangle \approx
280$ MeV/$c$).   At this missing momentum, none of the measured response
functions show a peak at $E_{\rm miss} \approx 40$ MeV where
single-particle knockout from the 1$s_{1/2}$-state is expected.
$R_{L+TT}$ is close to zero and the DWIA calculation. However,
$R_T$ and $R_{LT}$ are much larger than the DWIA calculation.  $R_T$
is also much larger than $R_{LT}$ indicating that the cross
section is due in large part to transverse two-body currents.  The fact that 
$R_{LT}$ is nonzero indicates that
$R_L$, although too small to measure directly, is also nonzero.

To summarize, we have measured  the  cross section and
response functions ($R_L, R_T$, and $R_{LT}$) for the $^{16}$O$(e,e^{\prime}p)$
 reaction  in quasielastic
kinematics at $Q^2 = 0.8$ (GeV/$c$)$^2$ and $\omega = 439$ MeV for
missing energies $25 < E_{\rm miss}< 120$ MeV at various missing momenta
$P_{\rm miss} \le 340$ MeV/$c$.  For $25 < E_{\rm miss} < 50$ MeV and 
$P_{\rm miss}
\approx 60$ MeV/$c$ the reaction is dominated by single-nucleon knockout
from the 1$s_{1/2}$-state and is described well by DWIA calculations.  
$(S_T - S_L)$ is smaller than that measured at
$Q^2 = 0.14$~\cite{ulmer1} and $Q^2 = 0.6$ (GeV/$c$)$^2$, but larger
than that measured at $Q^2 = 1.8$ (GeV/$c$)$^2$~\cite{dutta}.  This is
consistent with the previous observation that, at low
$P_{\rm miss}$,  knockout processes due to MEC and IC decrease with 
$Q^2$~\cite{dutta}.

At increasing  missing momenta, the importance of the  single-particle aspects 
is diminished.  The
cross section and the response functions no longer peak at the maximum
of the $s$-shell (40 MeV).  They no longer have the expected Lorentzian
shape for $s$-shell knockout.   DWIA calculations underestimate the cross 
section and response functions at
$P_{\rm miss} > 200$ MeV/$c$ by more than a factor of 10.  Hence, we conclude 
that the single-particle aspect of the 1$s_{1/2}$-state contributes less than 
10\% to the cross section at $P_{\rm miss} > 200$ MeV/$c$. This is in contrast 
to    the $p$-shell case, where DWIA calculations describe the data
well up to $P_{\rm miss} = 340$ MeV/$c$.

At $25 < E_{\rm miss} < 120$ and $P_{\rm miss} > 200$ MeV/$c$ the cross section
is almost constant in  missing energy and missing momentum. For 
$E_{\rm miss} > 60$ MeV this feature is well reproduced by two-nucleon knockout
 calculations, $(e,e^{\prime}pp)$ plus $(e,e^{\prime}pn)$. These calculations 
also account for the predominantly transverse nature of the cross section, due 
to the large contribution from the two-body (pion exchange and isobar) currents.
  This indicates  that the excess
continuum strength at high $P_{\rm miss}$ is dominated by two body
processes rather than by correlations. To our knowledge, this is the only model
 which can account for the shape, transversity and about the half of the 
magnitude of the measured continuum cross section. The unaccounted for 
strength suggests that additional currents and processes play an important role.

We acknowledge the outstanding support of the staff of the Accelerator
and Physics Divisions at Jefferson Laboratory that made this experiment
successful.  We thank Dr.~J.~Ryckebusch for providing theoretical
calculations.  We also thank Dr.~J.M.~Udias for providing us with the
NLSH bound-state wave functions.  This work was supported in part by
the U.S. Department of Energy contract DE-AC05-84ER40150
under which the Southeastern Universities Research Association
(SURA) operates the Thomas Jefferson National Accelerator Facility, other 
Department of Energy contracts, the National Science Foundation, the
Italian Istituto Nazionale di Fisica Nucleare (INFN), the French
Atomic Energy Commission and National Center of Scientific Research,
and the Natural Sciences and Engineering Research Council of Canada.

\begin{table}[H]
\begin{center}
\begin{tabular}{|@{\hspace{6.4mm}}c@{\hspace{6.4mm}}|@{\hspace{6.4mm}}c@{\hspace
{6.4mm}}
|@{\hspace{6.4mm}}c@{\hspace{6.4mm}}|}
$E_{\rm beam}$ & $\theta_e$ & $\theta_{pq}$ \\
(GeV) & ($^\circ$) & ($^\circ$) \\
\hline
0.843 & 100.7 & 0, 8, 16 \\
\hline
1.643 & 37.2 & 0, $\pm8$ \\
\hline
2.442 & 23.4 & 0, $\pm2.5$, $\pm8$, $\pm16$, $\pm20$ \\
\end{tabular}
\caption{Experimental Kinematics.}
\label{ExpKin}
\end{center}
\end{table}

\begin{figure}[H]
\centering \epsfig{file=fig3.epsi,width=6.5cm} 
\caption 
{Separated response functions for $\langle P_{\rm miss}\rangle \approx $ 
145 MeV/$c$.}
\label{responce_145} 
\end{figure}

\begin{figure}[H]
\centering \epsfig{file=fig4.epsi,width=6.5cm} 
\caption 
{Separated response functions for $\langle
P_{\rm miss}\rangle \approx$ 280 MeV/$c$.}
\label{responce_280} 
\end{figure}

\end{document}